\documentclass[twocolumn]{aastex631}

\usepackage[utf8]{inputenc}
\usepackage[T1]{fontenc}
\usepackage[normalem]{ulem}

\received{July 7, 2022}
\revised{xxx}
\submitjournal{ApJ}

\shorttitle{Radial Evolution}
\shortauthors{Halekas et al.}

\graphicspath{{./}{figures/}}

\begin{document}

\title{The Radial Evolution of the Solar Wind as Organized by Electron Distribution Parameters}

\correspondingauthor{Jasper S. Halekas}
\email{jasper-halekas@uiowa.edu}

\author[0000-0001-5258-6128]{J.~S. Halekas}
\affil{Department of Physics and Astronomy, 
University of Iowa, 
Iowa City, IA 52242, USA}

\author[0000-0002-7287-5098]{P. Whittlesey}
\affil{Space Sciences Laboratory, University of California, Berkeley, CA 94720, USA}

\author[0000-0001-5030-6030]{D.~E. Larson}
\affil{Space Sciences Laboratory, University of California, Berkeley, CA 94720, USA}

\author[0000-0001-6172-5062]{M. Maksimovic}
\affil{LESIA, Observatoire de Paris, Universite PSL, CNRS, Sorbonne Universite, Universite de Paris, 5 place Jules Janssen, 92195 Meudon, France}

\author[0000-0002-0396-0547]{R. Livi}
\affil{Space Sciences Laboratory, University of California, Berkeley, CA 94720, USA}

\author[0000-0001-6235-5382]{M. Berthomier}
\affil{Laboratoire de Physique des Plasmas, CNRS, Sorbonne Universite, Ecole Polytechnique, Observatoire de Paris, Universite Paris-Saclay, Paris, 75005, France}

\author[0000-0002-7077-930X]{J.~C. Kasper}
\affiliation{BWX Technologies, Inc., Washington DC 20002, USA}
\affiliation{Climate and Space Sciences and Engineering, University of Michigan, Ann Arbor, MI 48109, USA}

\author[0000-0002-3520-4041]{A.~W. Case}
\affil{Smithsonian Astrophysical Observatory, Cambridge, MA 02138, USA}

\author[0000-0002-7728-0085]{M.~L. Stevens}
\affil{Smithsonian Astrophysical Observatory, Cambridge, MA 02138, USA}

\author[0000-0002-1989-3596]{S.~D. Bale}
\affil{Space Sciences Laboratory, University of California, Berkeley, CA 94720, USA}
\affil{Physics Department, University of California, Berkeley, CA 94720, USA}

\author[0000-0003-3112-4201]{R.~J. MacDowall}
\affil{NASA/Goddard Space Flight Center, Greenbelt, MD 20771, USA}

\author[0000-0002-1573-7457]{M.~P. Pulupa}
\affil{Space Sciences Laboratory, University of California, Berkeley, CA 94720, USA}

\begin{abstract}
We utilize observations from the Parker Solar Probe (PSP) to study the radial evolution of the solar wind in the inner heliosphere. We analyze electron velocity distribution functions observed by the Solar Wind Electrons, Alphas, and Protons suite to estimate the coronal electron temperature and the local electric potential in the solar wind. From the latter value and the local flow speed, we compute the asymptotic solar wind speed. We group the PSP observations by asymptotic speed, and characterize the radial evolution of the wind speed, electron temperature, and electric potential within each group. In agreement with previous work, we find that the electron temperature (both local and coronal) and the electric potential are anti-correlated with wind speed. This implies that the electron thermal pressure and the associated electric field can provide more net acceleration in the slow wind than in the fast wind. We then utilize the inferred coronal temperature and the extrapolated electric + gravitational potential to show that both  electric field driven exospheric models and the equivalent thermally driven hydrodynamic models can explain the entire observed speed of the slowest solar wind streams. On the other hand, neither class of model can explain the observed speed of the faster solar wind streams, which thus require additional acceleration mechanisms. 
\end{abstract}


\section{Introduction} \label{sec:intro}

Despite decades of observational and theoretical study, the details of the mechanisms that accelerate the solar wind remain uncertain. The prediction of the existence of a supersonic solar wind by Parker \citep{parker_dynamics_1958} shortly before its discovery \citep{gringauz_study_1960, neugebauer_solar_1962} remains a triumph of theoretical physics. However, though Parker-type thermally driven wind scenarios can provide the needed acceleration to explain the slow solar wind, they have difficulty explaining the observed speeds of the fast solar wind, at least for realistic coronal temperatures \citep{parker_dynamical_1965, leer_acceleration_1982, hansteen_solar_2012}. Furthermore, observations of minor ion ratios indicate a strong anti-correlation between the speed of the solar wind and the coronal electron temperature \citep{geiss_southern_1995, gloeckler_implications_2003}, contrary to expectations for the purely thermally driven hydrodynamic models. This puzzle has stimulated the development of a variety of alternative wind acceleration models, many of which involve waves and/or magnetic reconnection \citep{hollweg_generation_2002, fisk_acceleration_2003, marsch_kinetic_2006, mccomas_understanding_2007, cranmer_self-consistent_2012}. Determining the relative importance of the various proposed acceleration mechanisms, and their roles in winds with different characteristics, represents a major focus for the currently operating Parker Solar Probe (PSP) \citep{fox_solar_2016} and Solar Orbiter \citep{muller_solar_2020} missions. 

So-called "exospheric" models \citep{jockers_solar_1970, lemaire_kinetic_1971, lemaire_kinetic_1973} provide an alternative framework for considering solar wind acceleration. These models, rather than utilizing a hydrodynamic (or magnetohydrodynamic) framework, leverage the requirement for global quasi-neutrality to find a self-consistent electric potential structure that results in a balance of both charge and current between the electrons and ions, given prescribed velocity distribution functions at the boundaries of the system. Given the much greater mobility of the electrons, the quasi-neutrality requirement naturally leads to the formation of an outward electric field in the solar wind that slows the escaping electrons and accelerates the ions. In the hydrodynamic limit, this electric field can be identified as the ambipolar electric field $- \nabla P_e/(e n_e)$. In fact, for Maxwellian electron distributions, and in the limit $m_e \rightarrow 0$, exospheric models become essentially equivalent to hydrodynamic models, as shown by \citet{parker_kinetic_2010}. Given this correspondence, exospheric models have the same basic difficulty in matching the observed speeds in the fast wind for realistic coronal temperatures. 

However, though exospheric models cannot easily account for collisional effects, they can straightforwardly incorporate the effects of non-Maxwellian velocity distributions \citep{scudder_causes_1992, pierrard_lorentzian_1996}. Exospheric models utilizing coronal electron velocity distribution functions (EVDFs) with a substantial non-thermal component can better reproduce the observed fast wind speeds \citep{maksimovic_kinetic_1997, zouganelis_transonic_2004}, as one would expect since an initial distribution with a greater suprathermal electron fraction requires a larger electric field to maintain quasi-neutrality. Unfortunately, it remains uncertain whether the required non-thermal EVDFs actually exist in the corona \citep{maksimovic_electron_2021}. Indeed, recent PSP measurements do not reveal substantial non-Maxwellian populations in the near-Sun EVDFs, and in fact actually suggest a decrease in the fractional abundance of suprathermal electrons close to the Sun \citep{bercic_coronal_2020, halekas_electrons_2020, abraham_radial_2022}. 

While the interplanetary electric field in the solar wind may not provide all of its acceleration, basic physical considerations indicate that it must provide some component of the acceleration. The only question is: How much? In this work, we utilize PSP observations to attempt to bound the answer to this question.

\section{Electron Velocity Distribution Functions and Parameters} \label{sec:vdfs}

We utilize data from the Solar Wind Electrons, Alphas, and Protons (SWEAP) instrument suite \citep{kasper_solar_2016} on PSP, specifically the Solar Probe ANalyzer-Ion (SPAN-Ion) sensor and the two Solar Probe ANalyzer-Electron (SPAN-Electron) \citep{whittlesey_solar_2020} sensors, to determine the local proton bulk speed and to characterize the EVDFs. We also utilize magnetometer data from the FIELDS suite \citep{bale_fields_2016} to organize the charged particle measurements. 

\begin{figure}
\plotone{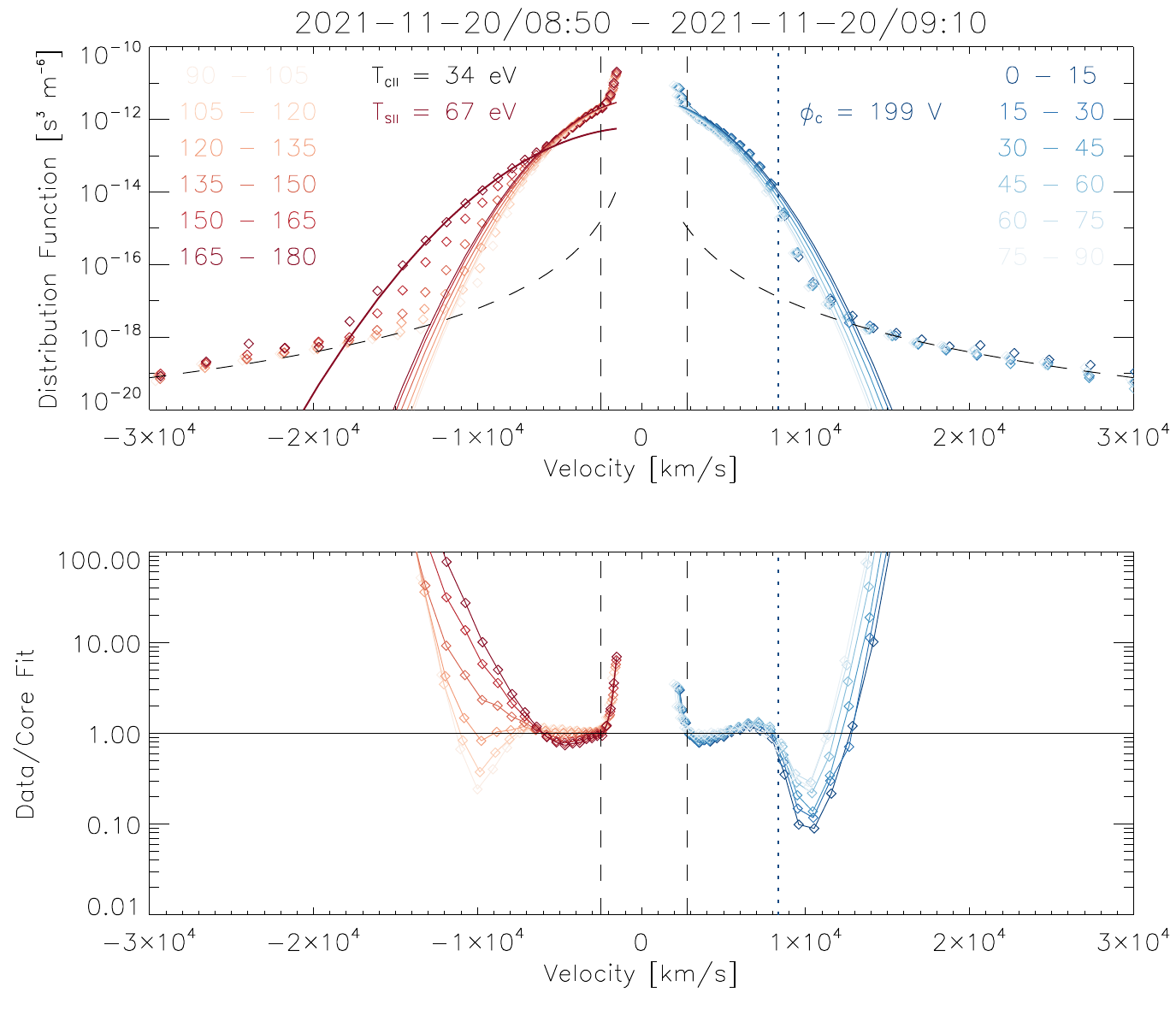}
\caption{Representative electron velocity distribution function (EVDF) observed in the near-Sun fast solar wind ($r \sim 18.5 \, R_S$, $v_p \sim 525  \textrm{ km/s}$). The top panel shows SPAN-Electron measurements binned by energy and pitch angle in the solar wind frame, accumulated over a 20-minute interval (colored diamonds). Positive velocity values represent pitch angles from $0-90^{\circ}$ (sunward-going electrons, given the sunward magnetic field at this time), and negative values represent pitch angles from $90-180^{\circ}$ (anti-sunward), as indicated. Colored curves in the top panel show the corresponding values for a drifting bi-Maxwellian fit to the core of the distribution and a Maxwellian fit to the suprathermal portion of the most anti-field aligned bin (which contains the strahl), with the corresponding parallel core and strahl temperatures ($T_{C||}$ and $T_{S||}$) indicated. The bottom panel shows the observed EVDF normalized by the core fit values. Vertical dashed lines in both panels delineate the low-energy portion of the EVDF contaminated by secondary electrons, and dashed curves in the top panel show the approximate instrumental background level. The vertical dotted line in both panels indicates the sunward cutoff in the EVDF, with the corresponding electric potential ($\phi_c$) indicated.  \label{fig:vdf1}}
\end{figure}

\begin{figure}
\plotone{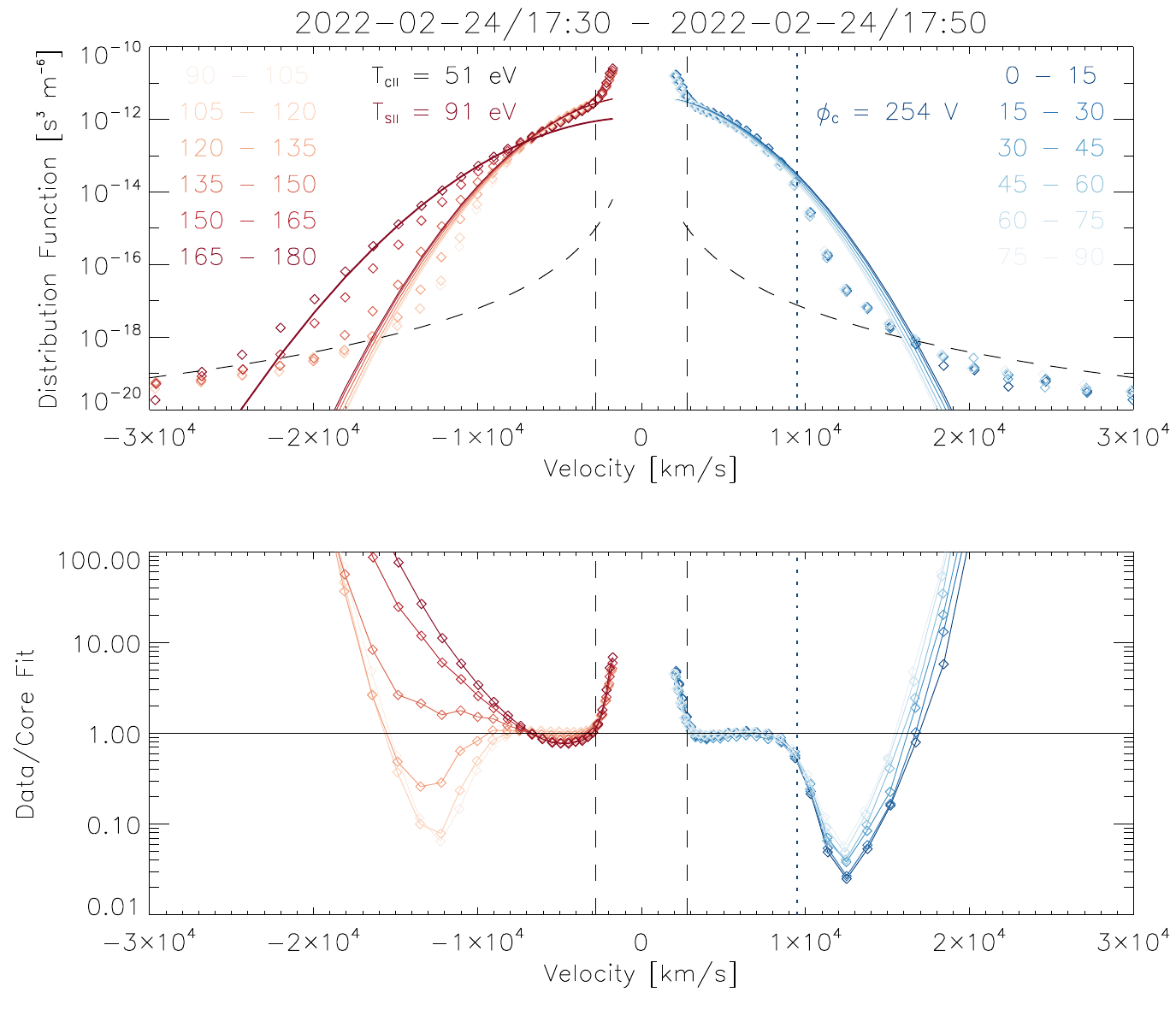}
\caption{Representative electron velocity distribution function observed in the near-Sun slow solar wind ($r \sim 18 \, R_S$, $v_p \sim 210 \textrm{ km/s}$), in the same format as Fig. \ref{fig:vdf1}.   \label{fig:vdf2}}
\end{figure}

To characterize each measured EVDF, we leverage analysis tools developed in a series of previous studies utilizing SPAN-Electron data \citep{bercic_coronal_2020, halekas_electrons_2020, halekas_sunward_2021}. In Fig. \ref{fig:vdf1} and Fig. \ref{fig:vdf2}, we show representative EVDFs observed in slow and fast solar wind streams at heliocentric distances of $18-18.5 \, R_S$, which serve to outline our methodology. We utilize proton bulk velocity moments from SPAN-Ion and magnetic field measurements from FIELDS to transform each measured EVDF to the proton frame and rotate it into field-aligned coordinates. In the remainder of this work we perform our analysis on each individual measured EVDF (obtained at cadences of 7-14 s). However, for these two illustrative examples we show data binned by pitch angle (in the plasma frame) and accumulated over longer time periods, in order to better illustrate the presence or absence of trace populations, and to show the instrumental background count level.

We first note the substantial similarity of the EVDFs observed in the slow and fast wind. Both have the same general structure, well known from observations at greater heliospheric distances, of a quasi-Maxwellian core at lower energies, and a suprathermal population at higher energies \citep{feldman_solar_1975, rosenbauer_survey_1977, pilipp_characteristics_1987, salem_precision_2021}. However, in contrast to measurements at greater heliocentric distances, the more collimated anti-sunward strahl component dominates the suprathermal population, with little to no halo present, as also shown by previous studies utilizing PSP observations \citep{halekas_electrons_2020, abraham_radial_2022}. Note that one recent study suggests that the halo may originate in the outer solar system \citep{horaites_heliospheric_2022}, which could explain this trend. In addition, near the Sun, the sunward portion of the EVDF actually falls below a Maxwellian function, revealing a "sunward deficit" \citep{halekas_electrons_2020, bercic_ambipolar_2021, halekas_sunward_2021}. 

 To retrieve the parameters of the electron core population, including the core parallel temperature $T_{C||}$, we fit the core portion of the EVDFs to drifting bi-Maxwellian functions, using the same methodology as \citet{halekas_electrons_2020}. As typically observed near the Sun \citep{halekas_electrons_2020, maksimovic_anticorrelation_2020}, we find a higher core temperature in the slower wind, and a lower core temperature in the faster wind. This difference in core temperature may result from the different initial conditions at the corona, consistent with the known wind speed-coronal temperature anti-correlation \citep{geiss_southern_1995, gloeckler_implications_2003}.  

We next analyze the suprathermal portion of the EVDFs to determine two critical parameters for our study. Utilizing the anti-sunward portion of the distribution, we employ a similar technique to \citet{bercic_coronal_2020} to find the effective parallel temperature of the strahl $T_{S||}$. To accomplish this, we fit the suprathermal portion of the EVDF in the most anti-field aligned bin (which contains the strahl) to the sum of a Maxwellian function representing the strahl and a second function representing the instrumental background (flat in units of count rate, and thus scaling with energy as $E^{-2}$ in units of distribution function), both of which we allow to vary in amplitude. \citet{bercic_coronal_2020} has shown observationally that the strahl parallel temperature does not vary appreciably with distance. Given that the strahl seemingly represents an escaping population from the corona, focused along the magnetic field line due to conservation of magnetic moment, this suggests that it can provide a remote measurement of the coronal electron temperature, as also supported by simulations \citep{bercic_interplay_2021}. In agreement with \citet{bercic_coronal_2020}, we find a higher strahl parallel temperature in the slow wind, and a lower strahl parallel temperature in the fast wind, with reasonable corresponding coronal temperatures of $0.75-1.1 \textrm{ MK}$. 

Finally, we analyze the sunward portion of the EVDF to determine the electron cutoff velocity. We employ the same technique as \citet{halekas_sunward_2021} to determine the cutoff velocity, dividing the measured EVDF by the core fit, and fitting to a hyperbolic tangent function to find the velocity where the EVDF drops to $50 \%$ of the core fit. The cutoff velocity and the associated sunward electron deficit \citep{halekas_electrons_2020, bercic_ambipolar_2021, halekas_sunward_2021, bercic_whistler_2021} arise as a natural consequence of the interplanetary electric field. This electric field, which accelerates the ions, forms a potential well for the electrons. Electrons with a kinetic energy smaller than the local potential $\phi_c$ (referenced to the value at infinity) remain trapped within the potential structure near the Sun, while those with larger kinetic energies escape to infinity. This forms a natural demarcation in the sunward suprathermal portion of the EVDF, which provides us with a clear marker of the local electric potential. In agreement with \citet{halekas_sunward_2021}, we find a larger potential in the slow solar wind, and a smaller potential in the fast solar wind, with values scaling approximately with the core temperature as $\phi_c \sim 5 \, T_{C||}$.

\section{Asymptotic Solar Wind Speed} \label{sec:asymptotic}

Any study of the radial evolution of the solar wind encounters the issue of aligning observations obtained at multiple heliocentric distances. Ideally speaking, one would wish to observe the same solar wind stream at a range of distances. However, practical realities make this nearly impossible, other than in special cases. Given a sufficiently large statistical sampling of data, one can define "wind families" at multiple radial distances, associate them to each other, and thereby attempt to trace the radial evolution of the wind \citep{maksimovic_anticorrelation_2020}. However, the PSP mission has arguably not yet achieved an adequate sampling of winds of various types (particularly fast winds) to attempt such a procedure. 

In this work, we instead attempt to classify the wind streams by their asymptotic speeds. To estimate the asymptotic speed, we make the ansatz that the only significant remaining acceleration and deceleration acting on the solar wind after its observation by PSP results from the gradients in the electric and gravitational potentials. We then can compute the asymptotic speed $v_{ASY}$ in terms of the local bulk speed $v_p$, electric potential $\phi_c$, and gravitational potential at heliocentric distance $r$.

\begin{equation}
v_{ASY} = \sqrt{\frac{2}{m_p} (\frac{1}{2} m_p {v_p}^2 + e \phi_c - \frac{G M_S m_p}{r})} \label{eq:vasy}
\end{equation}

This calculation of the asymptotic speed necessarily neglects any non-electric field wind acceleration mechanisms that act outside of the PSP orbit, as well as stream-stream interactions that can decelerate faster solar wind streams in the outer heliosphere. Therefore, we regard our calculation of the asymptotic speed only as an approximation; however, as long as the error terms remain comparatively small, it can still provide a meaningful first-order result. 

The solar wind near the Sun commonly contains a large number of brief short-lived increases in the local solar wind speed, associated with magnetic field rotations as expected for quasi-Alfv\'enic fluctuations, and thus often termed "switchbacks" \citep{bale_highly_2019, kasper_alfvenic_2019}. We wish to classify our measurements by the background speed rather than by the speed of these brief increases, and so we replace each local determination of the asymptotic speed by the median in a 20-minute interval. We perform the same procedure on the strahl parallel temperature values and the electric potentials derived from the sunward cutoff velocities in the work that follows. All other parameters employed in this study simply utilize the instantaneous local measurements. 

\begin{figure}
\plotone{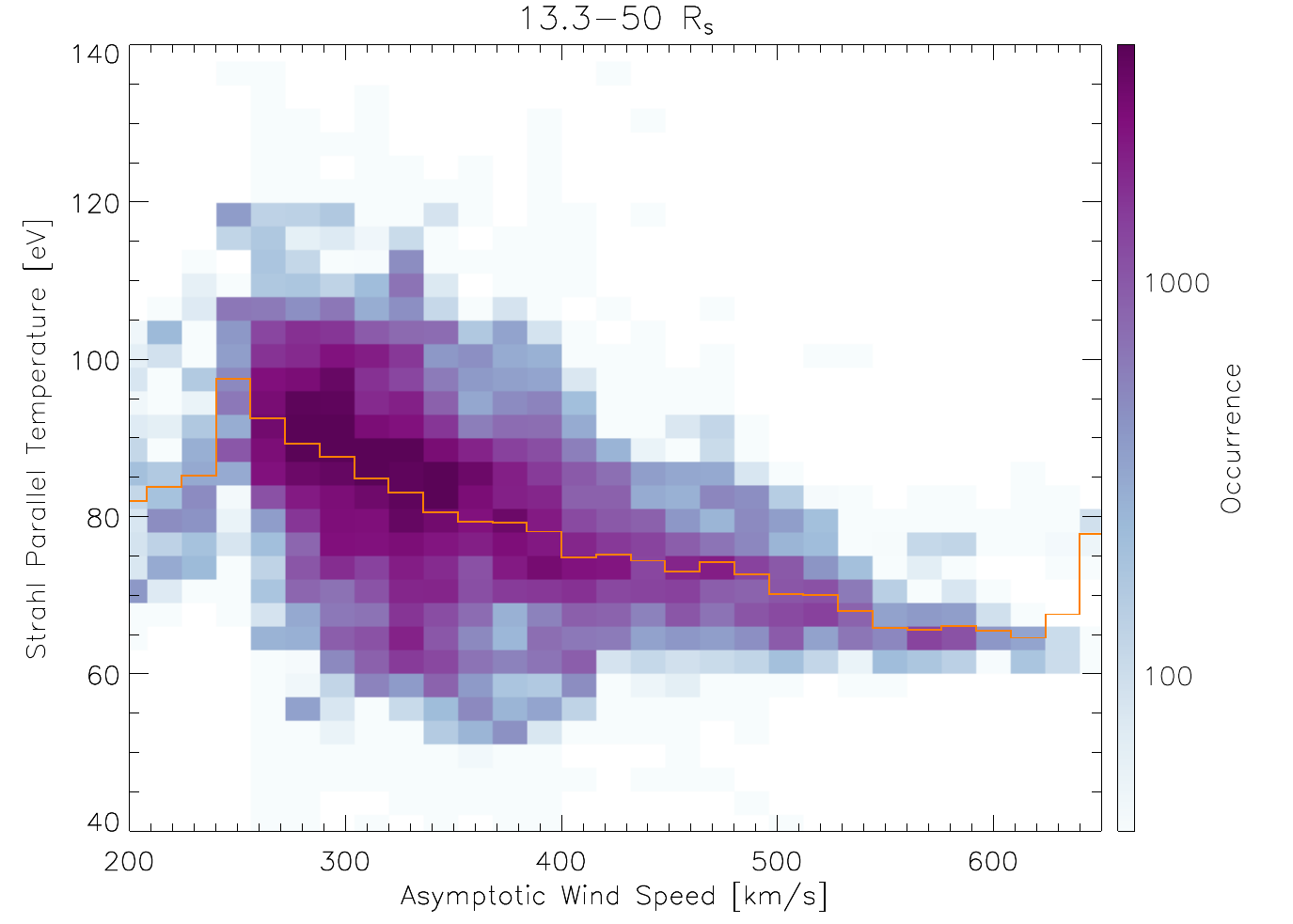}
\caption{Joint occurrence of asymptotic solar wind speed ($v_{ASY}$) and strahl parallel temperature ($T_{S||}$), determined as described in the text from PSP measurements from encounters 3-11 over heliocentric distances of $13.3-50 \, R_S$. The orange line shows the average value of the strahl parallel temperature for each speed bin. \label{fig:stp_vasy}}
\end{figure}

To assess the results of our calculation of asymptotic velocity and strahl parallel temperature, we show the occurrence of $(v_{ASY},T_{S||})$ pairs in our data set in Fig. \ref{fig:stp_vasy}. In agreement with \citet{bercic_coronal_2020}, we find a clear anti-correlation between strahl parallel temperature and wind speed, with very comparable numerical values (given the coverage in radius of \citet{bercic_coronal_2020}, we consider their local speed values nearly equivalent to our asymptotic speed values). The majority of the observations follow the same trend, with a range of extrapolated coronal temperatures of ${\sim} 95 \textrm{ eV} = 1.1 \textrm{ MK}$ for asymptotic speeds of ${\sim} 250 \textrm{ km/s}$ to ${\sim} 65 \textrm{ eV} = 0.75 \textrm{ MK}$ for asymptotic speeds of ${\sim} 600 \textrm{ km/s}$.

A small fraction of the data points, particularly at low estimated asymptotic speeds, have strahl parallel temperatures that lie below the prevailing trend. These could represent data points where the fitting methodology failed; however, targeted spot checks did not reveal any pervasive issues with the fitting procedure. Therefore, we postulate that these off-trend values may represent solar wind streams that still have significant acceleration remaining outside of the PSP location (i.e., observations for which our computation of the asymptotic speed would represent a significant underestimate). Most of these data points occur at low heliocentric distances, lending credence to this hypothesis. These solar wind streams may represent an appealing target for future analysis. However, in the remainder of this work we afford these points no special treatment.

\section{Radial Evolution of the Solar Wind} \label{sec:radial}

Given values of asymptotic speed (and strahl parallel temperature) computed as above for each data point, we can proceed to group our observations into bins and investigate the radial evolution of the solar wind within each bin. In the following analysis, we adopt a consistent binning strategy wherein we group the observations into 9 bins of asymptotic speed ranging from 200-650 km/s, each covering a 50 km/s range. 

\begin{figure}
\plotone{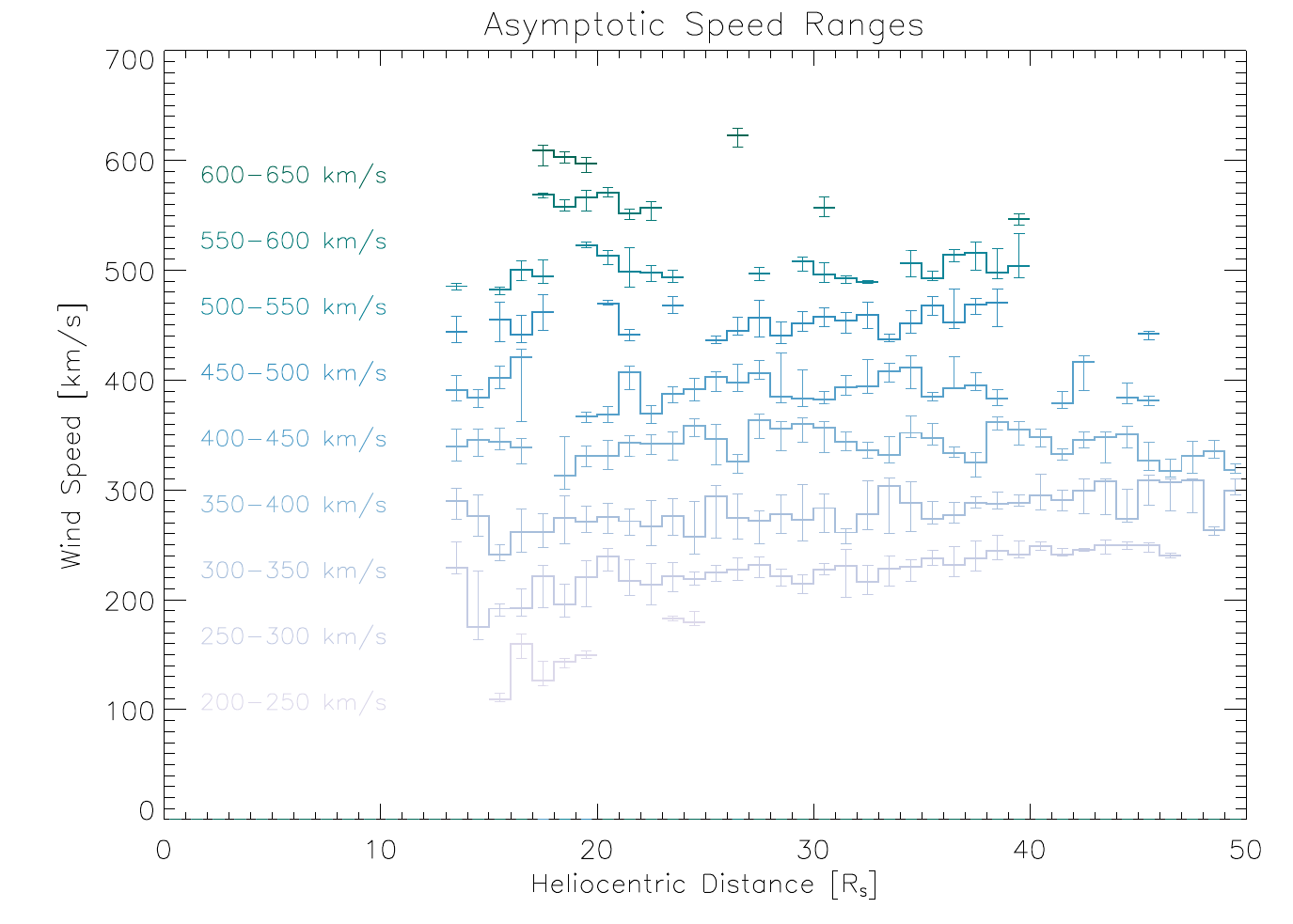}
\caption{Radial evolution of the local bulk solar wind proton speed ($v_p$) measured by SPAN-Ion, binned by asymptotic solar wind speed ($v_{ASY}$) in 9 ranges, as indicated. Colored lines represent median values for each radius-speed bin, and error bars represent upper and lower quartiles. \label{fig:vp_vasy}}
\end{figure}

In Fig. \ref{fig:vp_vasy}, we show the local proton speed binned in this manner, for the same data set as in Fig. \ref{fig:stp_vasy}. We find that the lower speed bins display a much larger acceleration than the upper speed bins, similar to previous results from \citet{maksimovic_anticorrelation_2020}. Our observations show that this trend results from two reinforcing factors. First, to achieve a given increase in speed, one requires a greater addition of energy (and thus a greater potential drop) for larger initial speeds, as a simple consequence of the quadratic dependence of kinetic energy on speed. Second, the observations reveal larger electric potential magnitudes in the slow solar wind, and thus the slower wind in the near-Sun environment will experience greater remaining acceleration. Given these two factors (and assuming our assumptions remain valid), most of the acceleration of the fast wind must occur below even the nearest PSP perihelion (to date) of $13.3 R_S$. On the other hand, a substantial fraction of the acceleration of the slow wind demonstrably occurs above this radius. 

\begin{figure}
\plotone{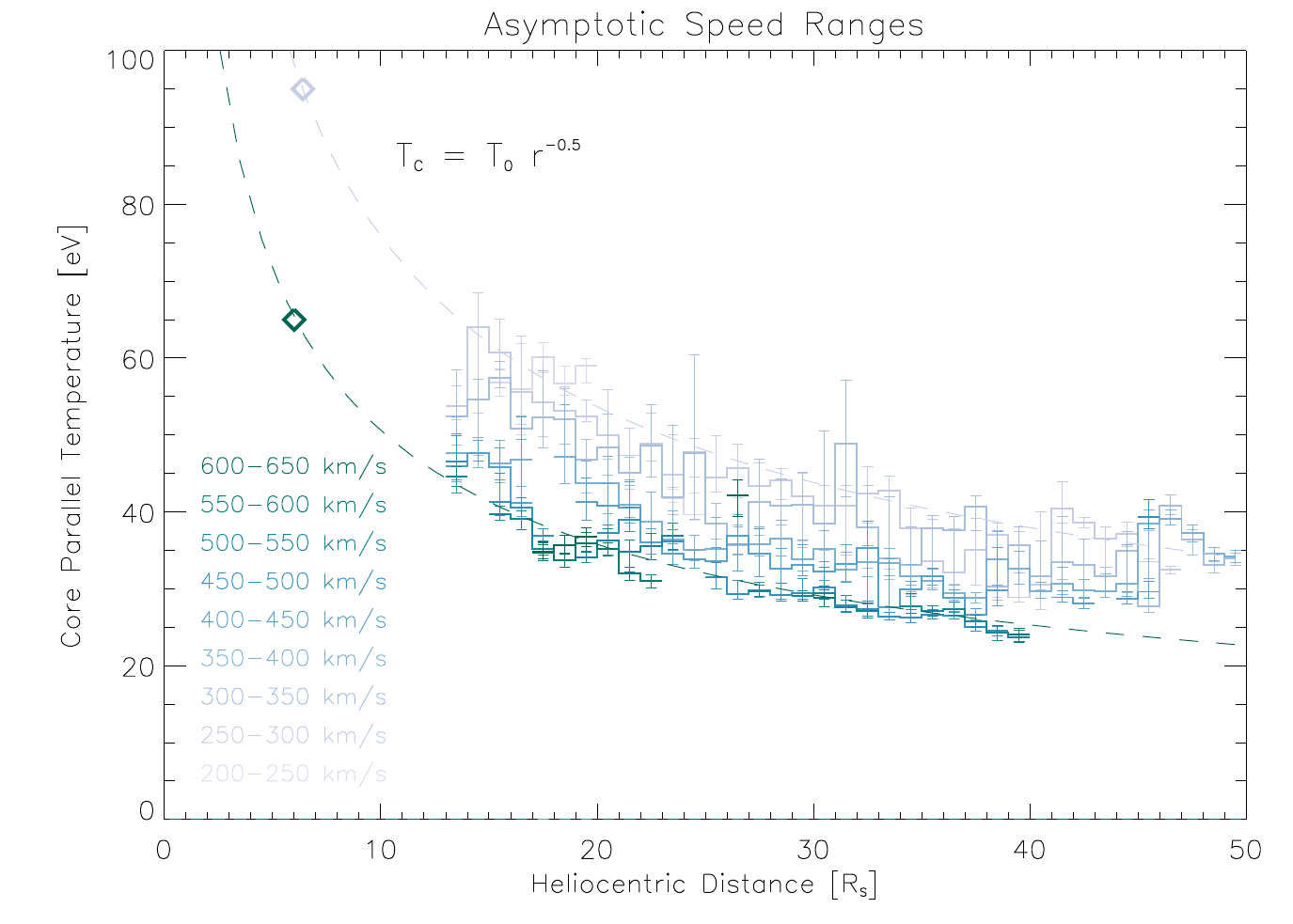}
\caption{Radial evolution of the local core parallel electron temperature ($T_{C||}$) measured by SPAN-Electron, binned by asymptotic solar wind speed ($v_{ASY}$) in 9 ranges, as indicated. Colored lines represent median values for each radius-speed bin, and error bars represent upper and lower quartiles. The two dashed curves show approximate power law extrapolations for low and high asymptotic speeds, with diamonds indicating the intersections of these curves with the corresponding strahl parallel temperature ($T_{S||}$) values.  \label{fig:tpar_vasy}}
\end{figure}

Binning by asymptotic speed provides a natural way to organize a number of other parameters of the solar wind. We choose just a few solar wind parameters to show in this work, using the same data set as in Fig. \ref{fig:stp_vasy} throughout. In Fig. \ref{fig:tpar_vasy} we show the core parallel electron temperature. As expected from previous work \citep{halekas_electrons_2020, maksimovic_anticorrelation_2020}, we find a clear anti-correlation between core parallel electron temperature and asymptotic speed, present at all heliocentric distances considered in this study.

As an illustrative exercise, in Fig. \ref{fig:tpar_vasy} we also show two power law curves that approximately represent the upper (low speed) and lower (high speed) envelopes of the observed core parallel temperatures. We do not derive the radial exponent $\alpha = -0.5$ from a rigorous fit or claim that exactly the same exponent holds for all asymptotic wind speeds, but merely choose this  as an approximately representative value. This exponent provides curves that correspond reasonably well with the data for all asymptotic speed bins. This radial exponent also appears roughly consistent with those derived in previous studies from observations at greater heliocentric distances \citep{marsch_cooling_1989, maksimovic_radial_2005}, and also with recent theoretical predictions \citep{boldyrev_electron_2020}. On the other hand, radial exponents consistent with purely adiabatic expansion ($\alpha = -4/3$) or with constant conductive luminosity ($\alpha = -2/7$) appear clearly incompatible with the observations. 

We also note the intersections of the power law extrapolations for the core parallel temperature with the corresponding values of the strahl parallel temperature. For both high and low asymptotic speeds, this occurs at ${\sim} 6 \, R_S$. If the power law extrapolations remain valid inside the closest PSP perihelion, and the strahl parallel temperature accurately records the coronal electron temperature, this would seemingly suggest that this radius approximately represents the source altitude of the solar wind (i.e. the exobase). However, our subsequent results in no way depend on this conclusion. 

\begin{figure}
\plotone{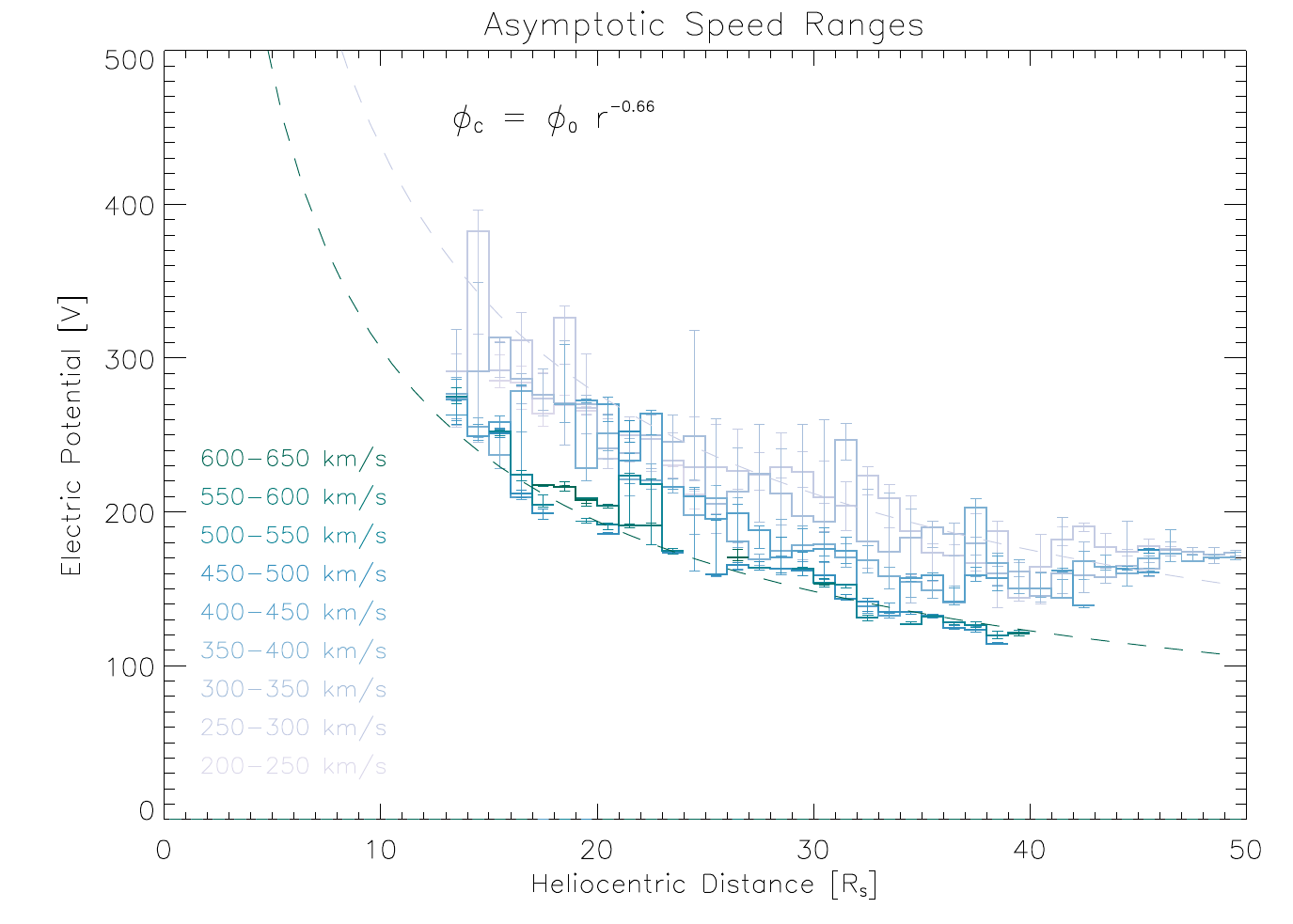}
\caption{Radial evolution of the local electric potential with respect to infinity ($\phi_c$), as determined from SPAN-Electron measurements of the sunward cutoff in the EVDF, binned by asymptotic solar wind speed ($v_{ASY}$) in 9 ranges, as indicated. Colored lines represent median values for each radius-speed bin, and error bars represent upper and lower quartiles. The two dashed curves show approximate power law extrapolations for low and high asymptotic speeds.  \label{fig:phic_vasy}}
\end{figure}

We next consider the radial evolution of the electric potential. In Fig. \ref{fig:phic_vasy}, we show the electric potential derived from the sunward cutoff in the EVDF as described above, again organized by asymptotic solar wind speed. Similar to the core parallel electron temperature, and again in accordance with previous work \citep{halekas_sunward_2021}, we find a clear anti-correlation between the magnitude of the electric potential and the asymptotic wind speed, present at all heliocentric distances considered in this study. This again indicates that the slower wind observed near the Sun has a larger remaining acceleration due to the electric potential than the faster wind.

We repeat the exercise of showing two power law curves that approximately represent the upper (low speed) and lower (high speed) envelopes of the observed electric potentials. We find that a slightly larger radial exponent, $\alpha = -0.66$, better represents the electric potential observations. \citet{bercic_ambipolar_2021} found the same radial exponent for the electric potential from PSP observations, and a slightly smaller exponent $\alpha = -0.55$ from simulations. Strictly speaking, for a power-law electron density and temperature, the corresponding ambipolar electric potential should have the same radial exponent as the temperature. The slight difference in exponents may indicate small deviations from power-law behavior, not particularly surprising given that significant wind acceleration occurs in this radial range. 

\begin{figure}
\plotone{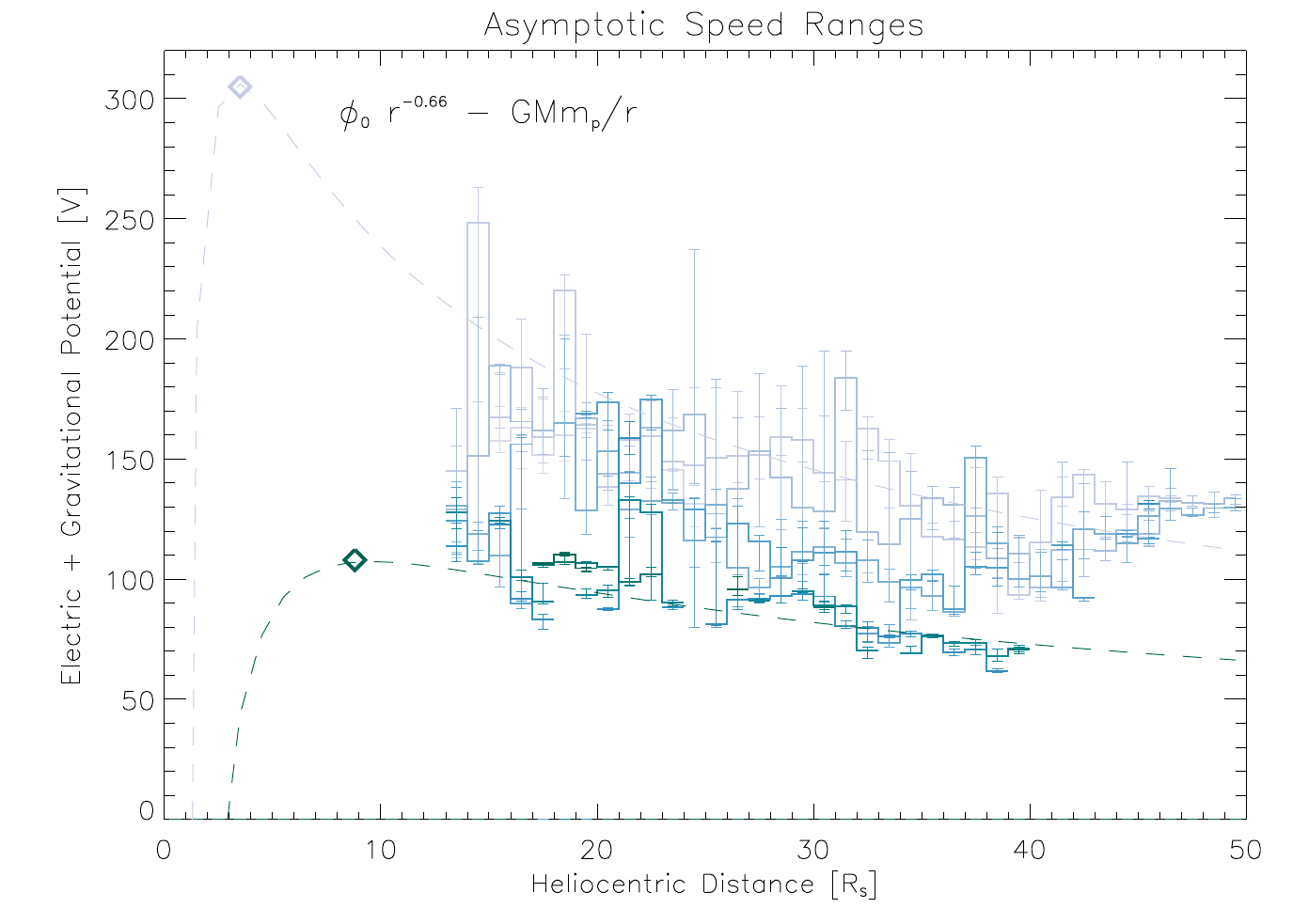}
\caption{Radial evolution of the sum of the electric potential (same as Fig. \ref{fig:phic_vasy}) and the gravitational potential per charge for protons, with respect to infinity, binned by asymptotic solar wind speed ($v_{ASY}$, same ranges as Fig. \ref{fig:phic_vasy}). Colored lines represent median values for each radius-speed bin, and error bars represent upper and lower quartiles. The two dashed curves show the same power law extrapolations as  Fig. \ref{fig:phic_vasy}, with the addition of the gravitational potential, with diamonds indicating the maxima in the total potential.  \label{fig:phitot_vasy}}
\end{figure}

For electrons, the gravitational potential remains largely inconsequential even at the smallest heliocentric distances. However, for ions, the gravitational potential becomes comparable to and even exceeds the electric potential at small heliocentric distances, since it varies as $r^{-1}$. Therefore, from the standpoint of solar wind acceleration, the sum of the two potentials represents a key quantity. We show this sum for the ions in Fig. \ref{fig:phitot_vasy}.

Given the relative magnitudes of the two potentials and the anti-correlation between electric potential and wind speed, the gravitational potential has a comparatively greater effect on the winds with faster asymptotic speeds. For these faster speed ranges, it almost cancels out the electric potential even at larger distances, and overtakes the extrapolated electric potential in magnitude at a comparatively large heliocentric radius of ${\sim} 9 \, R_S$. For the slower speed ranges, on the other hand, the gravitational potential does not overtake the extrapolated electric potential in magnitude until a heliocentric radius of ${\sim} 4 \, R_S$. This implies that the net potential acceleration of the slow wind can begin at a much closer distance, and that it has a much greater magnitude, in comparison to the fast wind. 

Meanwhile, ions inside of these critical radii experience a net inward rather than outward force due to the combined potentials. Of course, other sources of acceleration may (or must) exist close to the Sun, and their influence would add to the potential acceleration or deceleration, possibly affecting the direction of the net force at some distances.

\section{Acceleration of the Solar Wind} \label{sec:acceleration}

As a final exercise, we consider the implications of our observations for the acceleration of the solar wind. Given the potential curves shown in Fig. \ref{fig:phitot_vasy}, and an estimate for the initial wind speed, we can derive the expected radial profiles of wind speed, given only potential acceleration. For the initial wind speed values, we choose the ion sound speeds corresponding to the coronal temperatures inferred from the relevant strahl parallel temperatures (i.e. $v_0 = u_s = \sqrt{2 k T/m_p}$). Thus, unlike \citet{bercic_ambipolar_2021}, we assume slightly different initial speeds for the slow and fast wind; however, the resulting speeds of 135 km/s and 112 km/s in the slow and fast wind respectively both correspond fairly closely with the initial speed of 121 km/s utilized in that previous work. Only the existence of proton temperatures very different from the electron temperatures would significantly change these initial speeds. More sophisticated models exist that take into account the possibility of a range of initial speeds and altitudes (i.e. multi-base models \citep{Martinovic:prep}), but we do not consider such effects in this simple exercise. 

\begin{figure}
\plotone{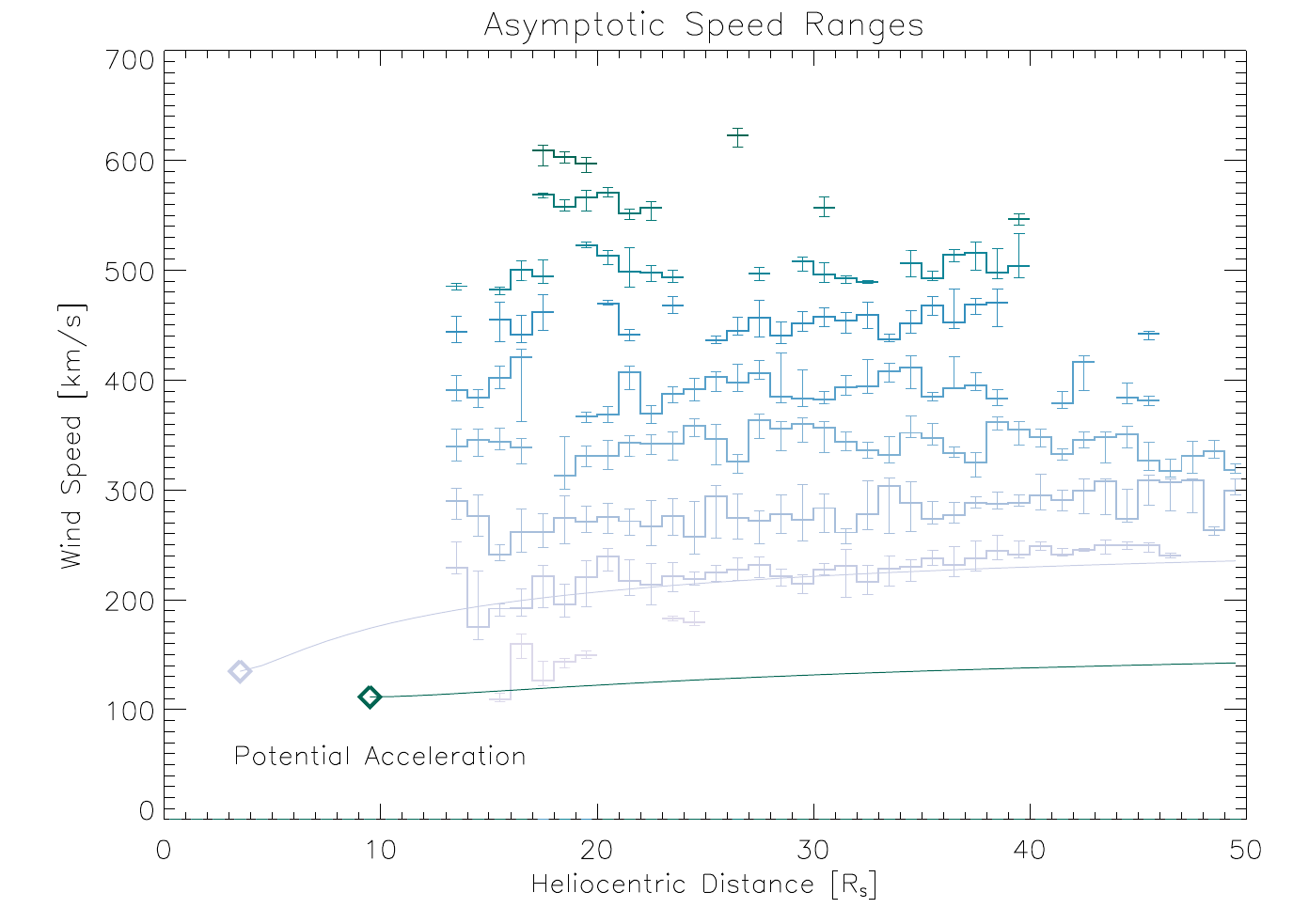}
\caption{Radial evolution of the local solar wind proton speed ($v_p$) measured by SPAN-Ion (same as Fig. \ref{fig:vp_vasy}), with curves showing the predicted speed profiles corresponding to the total potential curves from Fig. \ref{fig:phitot_vasy}, with initial values estimated from the strahl parallel temperatures using $v_0 = \sqrt{2 k T_{S||}/m_p}$.  \label{fig:exov_vasy}}
\end{figure}

Given initial speed estimates, we can simply utilize conservation of energy to derive the expected radial speed profiles corresponding to the total potential profiles for slow and fast winds. We show the results in Fig. \ref{fig:exov_vasy}. We find that the predicted potential acceleration curve for the slow wind matches the observations for the lower asymptotic speed bins rather well. On the other hand, the predicted acceleration curve for the fast wind does not correspond well with the observations.

One should not make too much of the fact that the slopes of the predicted speed curves match those of the observations. Since we organized the observations into asymptotic speed ranges utilizing the local values of the electric and gravitational potentials (and conservation of energy), the radial dependence of the total potential for each asymptotic speed range is already in essence built into our analysis. A significant difference in the slopes would represent an inconsistency that would require revisiting our assumptions. So, the agreement in the slopes merely verifies the self-consistency of our assumptions.  However, our analysis does clearly show that, for any reasonable initial conditions, the net potential acceleration cannot explain the observed fast wind speeds, or even a significant fraction thereof. On the other hand, for very reasonable initial conditions, the net potential acceleration can explain the entire observed acceleration of the slowest ranges of solar wind. 

We compare our results to two recent previous studies. \citet{bercic_ambipolar_2021} analyzed a set of PSP data representing a mix of wind speeds, and found that electrostatic acceleration could explain $77 \%$ of the observed speed of the solar wind (related simulations suggest an even higher percentage \citep{bercic_interplay_2021}). The results in Fig. 7 of \citet{bercic_ambipolar_2021} show that the observed speed values one standard deviation below the mean nearly follow the predicted acceleration curve. This appears entirely consistent with our result that potential acceleration can fully explain the lowest range of asymptotic wind speeds.  

Meanwhile, \citet{horaites_heliospheric_2022} analyzed PSP data using a Liouville method, and found that the potential thereby inferred could explain nearly $100 \%$ of the solar wind acceleration over a radial range of 0.18-0.79 AU. Though this radial range barely overlaps that considered in this study, the results agree with those from this work. Furthermore, these previous results also support the validity of our initial assumption that only potential acceleration occurs outside of the location of the PSP observations used in this study. 

As an additional point of comparison, we repeat our exercise for a thermally driven hydrodynamic model. We utilize a basic Parker model \citep{parker_hydrodynamic_1960}, derived for an empirically chosen polytropic index $\gamma = 5/4$. For an $r^{-2}$ density radial dependence (not precisely accurate given that the wind experiences a net acceleration in the radial range of interest, but still a reasonable first approximation), this polytropic index leads to a temperature radial exponent $\alpha = -0.5$, approximately consistent with the PSP observations (see Fig. \ref{fig:tpar_vasy}). For initial coronal temperature estimates, we again utilize the strahl parallel temperatures corresponding to the slow and fast asymptotic speed ranges, as above. The existence of proton temperatures very different from the electron temperatures, or of very different electron and proton polytropic indices, would alter the initial speeds and acceleration profiles (see \citet{Dakeyo:prep} for a recent model that incorporates both such effects). 

\begin{figure}
\plotone{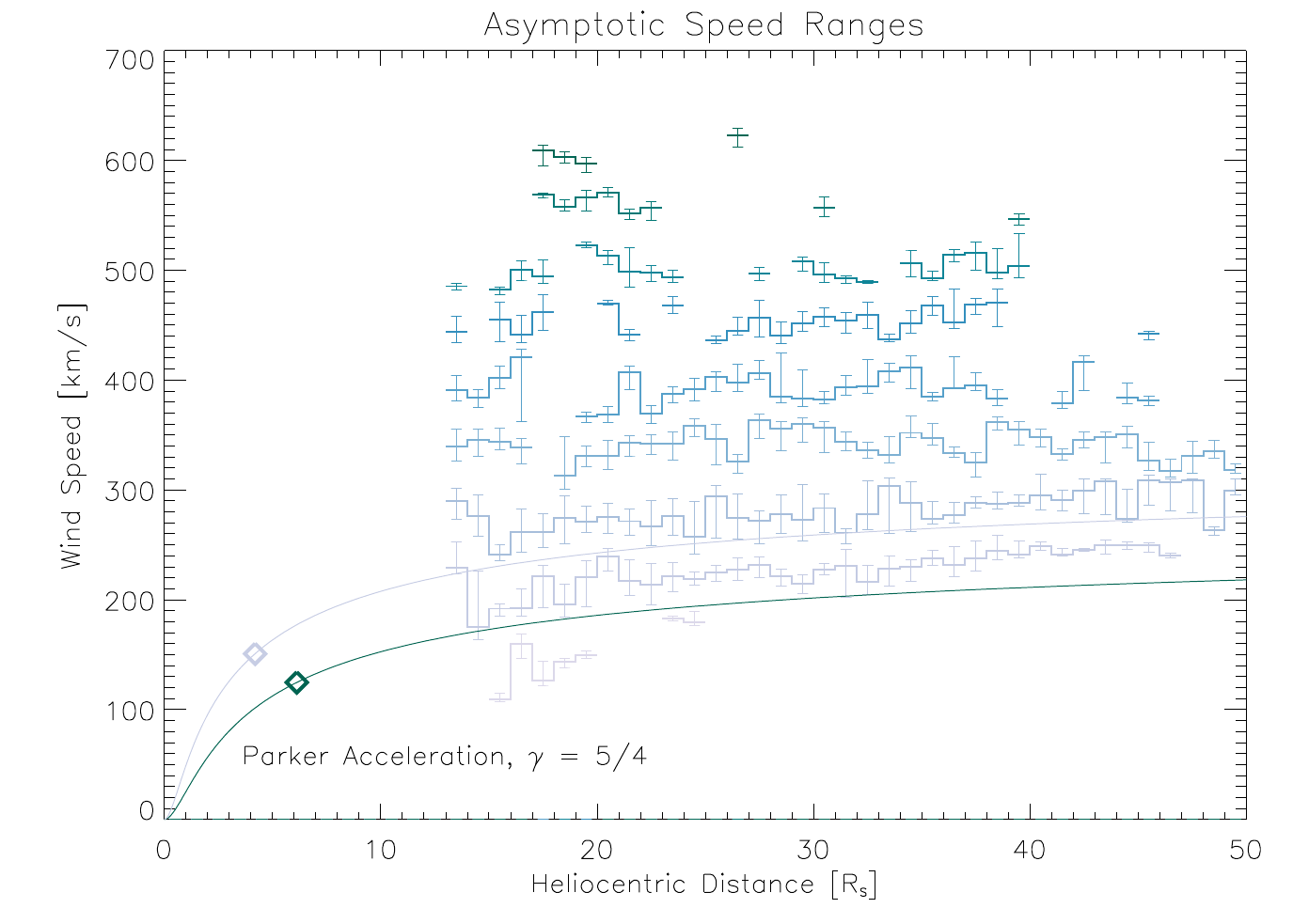}
\caption{Radial evolution of the local solar wind proton speed ($v_p$) measured by SPAN-Ion (same as Figs. \ref{fig:vp_vasy} and \ref{fig:exov_vasy}), with curves showing the predicted speed profiles from Parker solar wind models for a polytropic index $\gamma = 5/4$ \citep{parker_hydrodynamic_1960}, with coronal temperatures set to the corresponding strahl parallel temperatures ($T_{S||}$). The diamonds show the critical points for the two solutions.  \label{fig:parkerv_vasy}}
\end{figure}

We show the resulting predicted wind speed curves in Fig. \ref{fig:parkerv_vasy}. As expected, the results appear very comparable to those for the potential model shown in Fig. \ref{fig:exov_vasy}. We note that the initial speeds at the critical point are once again the ion sound speeds (as for all Parker models), but now given by $v_0 = u_s = \sqrt{2 \gamma k T/m_p}$, so slightly greater ($10 \%$) than the assumed initial speeds for the potential model. The net predicted acceleration for the Parker model agrees rather closely with that from the potential model for the slow wind case, while the Parker model provides a slightly larger acceleration for the fast wind case.

Though a few small differences exist, we can draw the same basic conclusions for both classes of models. Either a potential model or a Parker model can explain the entire acceleration of the slowest solar wind, given very reasonable initial conditions. On the other hand, for any reasonable initial conditions, neither the potential model nor the Parker model can explain the net acceleration of the faster solar wind streams.

\section{Conclusions and Implications} \label{sec:conc}

None of the conclusions of this study should come as a surprise. The basic difficulties in achieving the speed of the fast wind through either a thermally driven hydrodynamic scenario or an exospheric model have been well understood for decades. However, this work provides new observationally based constraints on how much of the acceleration other mechanisms must provide, and in what radial ranges. Our results indicate that, for the slow solar wind, we require no acceleration mechanisms other than the electric potential to explain the PSP observations. On the other hand, for the faster wind streams, some additional mechanism(s) must provide many hundreds of km/s of acceleration, rather close to the Sun, in order to explain the observations. 

One obvious way to rescue the exospheric solar wind models in the fast wind case would be to postulate the presence of a significant suprathermal electron population in the corona at the source of the fast solar wind. We note that the presence of such a population becomes less and less likely the closer that we come to the Sun without measuring it directly. The observations shown in Fig. \ref{fig:vdf1} and Fig. \ref{fig:vdf2} suggest that, if such a population exists, it must have very low values of distribution function in comparison to the core and strahl to escape our observational capability. Even with a fairly long integration, no significant suprathermal component appears clearly above background in either the slow or fast wind EVDFs. A more detailed analysis could potentially address the question of whether a suprathermal component both capable of providing the required additional acceleration and compatible with the in situ near-Sun observations could still exist.

From a more "glass half full" point of view, one can regard the current results as at least a partial triumph for both the exospheric models and the appropriate corresponding hydrodynamic models. Our results indicate that the electron pressure gradient and the associated electric potential can provide the entire acceleration required to explain the slowest solar wind streams. This implies that no additional acceleration mechanisms need operate in the slow wind. Of course, this does not rule out the operation of additional mechanisms, but it implies that they must not result in net acceleration significant in comparison to the electric field, in the slow wind.

\begin{acknowledgments}
This work was supported by the Parker Solar Probe (PSP) mission and the Solar Wind Electrons, Alphas, and Protons (SWEAP) team through contract NNN06AA01C. PSP was designed, built, and is now operated by the Johns Hopkins Applied Physics Laboratory as part of NASA’s Living with a Star (LWS) program. Support from the LWS management and technical team has played a critical role in the success of the PSP mission. We acknowledge insightful discussions within the ``Heliospheric Energy Budget: From Kinetic Scales to Global Solar Wind Dynamics'' team led by M.~E. Innocenti and A.~Tenerani at the International Space Science Institute (ISSI).
\end{acknowledgments}

\bibliography{references, prep}

\begin{thebibliography}{}
\expandafter\ifx\csname natexlab\endcsname\relax\def\natexlab#1{#1}\fi
\providecommand{\url}[1]{\href{#1}{#1}}
\providecommand{\dodoi}[1]{doi:~\href{http://doi.org/#1}{\nolinkurl{#1}}}
\providecommand{\doeprint}[1]{\href{http://ascl.net/#1}{\nolinkurl{http://ascl.net/#1}}}
\providecommand{\doarXiv}[1]{\href{https://arxiv.org/abs/#1}{\nolinkurl{https://arxiv.org/abs/#1}}}

\bibitem[{Abraham {et~al.}(2022)Abraham, Owen, Verscharen, Bakrania, Stansby,
  Wicks, Nicolaou, Whittlesey, Rueda, Jeong, \& Berčič}]{abraham_radial_2022}
Abraham, J.~B., Owen, C.~J., Verscharen, D., {et~al.} 2022, The Astrophysical
  Journal, 931, 118, \dodoi{10.3847/1538-4357/ac6605}

\bibitem[{Bale {et~al.}(2016)Bale, Goetz, Harvey, Turin, Bonnell,
  Dudok de Wit, Ergun, MacDowall, Pulupa, Andre, Bolton, Bougeret, Bowen,
  Burgess, Cattell, Chandran, Chaston, Chen, Choi, Connerney, Cranmer,
  Diaz-Aguado, Donakowski, Drake, Farrell, Fergeau, Fermin, Fischer, Fox,
  Glaser, Goldstein, Gordon, Hanson, Harris, Hayes, Hinze, Hollweg, Horbury,
  Howard, Hoxie, Jannet, Karlsson, Kasper, Kellogg, Kien, Klimchuk,
  Krasnoselskikh, Krucker, Lynch, Maksimovic, Malaspina, Marker, Martin,
  Martinez-Oliveros, McCauley, McComas, McDonald, Meyer-Vernet, Moncuquet,
  Monson, Mozer, Murphy, Odom, Oliverson, Olson, Parker, Pankow, Phan,
  Quataert, Quinn, Ruplin, Salem, Seitz, Sheppard, Siy, Stevens, Summers,
  Szabo, Timofeeva, Vaivads, Velli, Yehle, Werthimer, \&
  Wygant}]{bale_fields_2016}
Bale, S.~D., Goetz, K., Harvey, P.~R., {et~al.} 2016, Space Science Reviews,
  204, 49, \dodoi{10.1007/s11214-016-0244-5}

\bibitem[{Bale {et~al.}(2019)Bale, Badman, Bonnell, Bowen, Burgess, Case,
  Cattell, Chandran, Chaston, Chen, Drake, de~Wit, Eastwood, Ergun, Farrell,
  Fong, Goetz, Goldstein, Goodrich, Harvey, Horbury, Howes, Kasper, Kellogg,
  Klimchuk, Korreck, Krasnoselskikh, Krucker, Laker, Larson, MacDowall,
  Maksimovic, Malaspina, Martinez-Oliveros, McComas, Meyer-Vernet, Moncuquet,
  Mozer, Phan, Pulupa, Raouafi, Salem, Stansby, Stevens, Szabo, Velli, Woolley,
  \& Wygant}]{bale_highly_2019}
Bale, S.~D., Badman, S.~T., Bonnell, J.~W., {et~al.} 2019, Nature, 576, 237,
  \dodoi{10.1038/s41586-019-1818-7}

\bibitem[{Berčič {et~al.}(2021{\natexlab{a}})Berčič, Landi, \&
  Maksimović}]{bercic_interplay_2021}
Berčič, L., Landi, S., \& Maksimović, M. 2021{\natexlab{a}}, Journal of
  Geophysical Research: Space Physics, 126, e2020JA028864,
  \dodoi{https://doi.org/10.1029/2020JA028864}

\bibitem[{Berčič {et~al.}(2020)Berčič, Larson, Whittlesey, Maksimović,
  Badman, Landi, Matteini, Bale, Bonnell, Case, Wit, Goetz, Harvey, Kasper,
  Korreck, Livi, MacDowall, Malaspina, Pulupa, \&
  Stevens}]{bercic_coronal_2020}
Berčič, L., Larson, D., Whittlesey, P., {et~al.} 2020, The Astrophysical
  Journal, 892, 88, \dodoi{10.3847/1538-4357/ab7b7a}

\bibitem[{Berčič {et~al.}(2021{\natexlab{b}})Berčič, Maksimović, Halekas,
  Landi, Owen, Verscharen, Larson, Whittlesey, Badman, Bale, Case, Goetz,
  Harvey, Kasper, Korreck, Livi, MacDowall, Malaspina, Pulupa, \&
  Stevens}]{bercic_ambipolar_2021}
Berčič, L., Maksimović, M., Halekas, J.~S., {et~al.} 2021{\natexlab{b}}, The
  Astrophysical Journal, 921, 83, \dodoi{10.3847/1538-4357/ac1f1c}

\bibitem[{Berčič {et~al.}(2021{\natexlab{c}})Berčič, Verscharen, Owen,
  Colomban, Kretzschmar, Chust, Maksimovic, Kataria, Anekallu, Behar,
  Berthomier, Bruno, Fortunato, Kelly, Khotyaintsev, Lewis, Livi, Louarn, Mele,
  Nicolaou, Watson, \& Wicks}]{bercic_whistler_2021}
Berčič, L., Verscharen, D., Owen, C.~J., {et~al.} 2021{\natexlab{c}},
  Astronomy \& Astrophysics, 656, A31, \dodoi{10.1051/0004-6361/202140970}

\bibitem[{Boldyrev {et~al.}(2020)Boldyrev, Forest, \&
  Egedal}]{boldyrev_electron_2020}
Boldyrev, S., Forest, C., \& Egedal, J. 2020, Proceedings of the National
  Academy of Sciences, 117, 9232, \dodoi{10.1073/pnas.1917905117}

\bibitem[{Cranmer(2012)}]{cranmer_self-consistent_2012}
Cranmer, S.~R. 2012, Space Science Reviews, 172, 145,
  \dodoi{10.1007/s11214-010-9674-7}

\bibitem[{Dakeyo {et~al.}(2022)}]{Dakeyo:prep}
Dakeyo, J.-B., {et~al.} 2022, The Astrophysical Journal, in preparation

\bibitem[{Feldman {et~al.}(1975)Feldman, Asbridge, Bame, Montgomery, \&
  Gary}]{feldman_solar_1975}
Feldman, W.~C., Asbridge, J.~R., Bame, S.~J., Montgomery, M.~D., \& Gary, S.~P.
  1975, Journal of Geophysical Research, 80, 4181,
  \dodoi{10.1029/JA080i031p04181}

\bibitem[{Fisk(2003)}]{fisk_acceleration_2003}
Fisk, L.~A. 2003, Journal of Geophysical Research: Space Physics, 108,
  \dodoi{10.1029/2002JA009284}

\bibitem[{Fox {et~al.}(2016)Fox, Velli, Bale, Decker, Driesman, Howard, Kasper,
  Kinnison, Kusterer, Lario, Lockwood, McComas, Raouafi, \&
  Szabo}]{fox_solar_2016}
Fox, N.~J., Velli, M.~C., Bale, S.~D., {et~al.} 2016, Space Science Reviews,
  204, 7, \dodoi{10.1007/s11214-015-0211-6}

\bibitem[{Geiss {et~al.}(1995)Geiss, Gloeckler, Steiger, Balsiger, Fisk,
  Galvin, Ipavich, Livi, McKenzie, Ogilvie, \& Et}]{geiss_southern_1995}
Geiss, J., Gloeckler, G., Steiger, R.~v., {et~al.} 1995, Science, 268, 1033,
  \dodoi{10.1126/science.7754380}

\bibitem[{Gloeckler {et~al.}(2003)Gloeckler, Zurbuchen, \&
  Geiss}]{gloeckler_implications_2003}
Gloeckler, G., Zurbuchen, T.~H., \& Geiss, J. 2003, Journal of Geophysical
  Research: Space Physics, 108, \dodoi{10.1029/2002JA009286}

\bibitem[{Gringauz {et~al.}(1960)Gringauz, Bezrokikh, Ozerov, \&
  Rybchinskii}]{gringauz_study_1960}
Gringauz, K.~I., Bezrokikh, V.~V., Ozerov, V.~D., \& Rybchinskii, R.~E. 1960,
  Soviet Physics Doklady, 5, 361

\bibitem[{Halekas {et~al.}(2020)Halekas, Whittlesey, Larson, McGinnis,
  Maksimovic, Berthomier, Kasper, Case, Korreck, Stevens, Klein, Bale,
  MacDowall, Pulupa, Malaspina, Goetz, \& Harvey}]{halekas_electrons_2020}
Halekas, J.~S., Whittlesey, P., Larson, D.~E., {et~al.} 2020, The Astrophysical
  Journal Supplement Series, 246, 22, \dodoi{10.3847/1538-4365/ab4cec}

\bibitem[{Halekas {et~al.}(2021)Halekas, Berčič, Whittlesey, Larson, Livi,
  Berthomier, Kasper, Case, Stevens, Bale, MacDowall, \&
  Pulupa}]{halekas_sunward_2021}
Halekas, J.~S., Berčič, L., Whittlesey, P., {et~al.} 2021, The Astrophysical
  Journal, 916, 16, \dodoi{10.3847/1538-4357/ac096e}

\bibitem[{Hansteen \& Velli(2012)}]{hansteen_solar_2012}
Hansteen, V.~H., \& Velli, M. 2012, Space Science Reviews, 172, 89,
  \dodoi{10.1007/s11214-012-9887-z}

\bibitem[{Hollweg \& Isenberg(2002)}]{hollweg_generation_2002}
Hollweg, J.~V., \& Isenberg, P.~A. 2002, Journal of Geophysical Research: Space
  Physics, 107, SSH 12, \dodoi{10.1029/2001JA000270}

\bibitem[{Horaites \& Boldyrev(2022)}]{horaites_heliospheric_2022}
Horaites, K., \& Boldyrev, S. 2022, The {Heliospheric} {Ambipolar} {Potential}
  {Inferred} from {Sunward}-{Propagating} {Halo} {Electrons}, Tech. Rep.
  arXiv:2204.06532, arXiv, \dodoi{10.48550/arXiv.2204.06532}

\bibitem[{Jockers(1970)}]{jockers_solar_1970}
Jockers, K. 1970, Astronomy and Astrophysics, 6, 219

\bibitem[{Kasper {et~al.}(2016)Kasper, Abiad, Austin, Balat-Pichelin, Bale,
  Belcher, Berg, Bergner, Berthomier, \& Bookbinder}]{kasper_solar_2016}
Kasper, J.~C., Abiad, R., Austin, G., {et~al.} 2016, Space Science Reviews,
  204, 131, \dodoi{10.1007/s11214-015-0206-3}

\bibitem[{Kasper {et~al.}(2019)Kasper, Bale, Belcher, Berthomier, Case,
  Chandran, Curtis, Gallagher, Gary, Golub, Halekas, Ho, Horbury, Hu, Huang,
  Klein, Korreck, Larson, Livi, Maruca, Lavraud, Louarn, Maksimovic,
  Martinovic, McGinnis, Pogorelov, Richardson, Skoug, Steinberg, Stevens,
  Szabo, Velli, Whittlesey, Wright, Zank, MacDowall, McComas, McNutt, Pulupa,
  Raouafi, \& Schwadron}]{kasper_alfvenic_2019}
Kasper, J.~C., Bale, S.~D., Belcher, J.~W., {et~al.} 2019, Nature, 576, 228,
  \dodoi{10.1038/s41586-019-1813-z}

\bibitem[{Leer {et~al.}(1982)Leer, Holzer, \& Flå}]{leer_acceleration_1982}
Leer, E., Holzer, T.~E., \& Flå, T. 1982, Space Science Reviews, 33, 161,
  \dodoi{10.1007/BF00213253}

\bibitem[{Lemaire \& Scherer(1971)}]{lemaire_kinetic_1971}
Lemaire, J., \& Scherer, M. 1971, Journal of Geophysical Research, 76, 7479,
  \dodoi{10.1029/JA076i031p07479}

\bibitem[{Lemaire \& Scherer(1973)}]{lemaire_kinetic_1973}
---. 1973, Reviews of Geophysics, 11, 427, \dodoi{10.1029/RG011i002p00427}

\bibitem[{Maksimovic {et~al.}(1997)Maksimovic, Pierrard, \&
  Lemaire}]{maksimovic_kinetic_1997}
Maksimovic, M., Pierrard, V., \& Lemaire, J.~F. 1997, Astronomy and
  Astrophysics, 324, 725

\bibitem[{Maksimovic {et~al.}(2021)Maksimovic, Walsh, Pierrard, Štverák, \&
  Zouganelis}]{maksimovic_electron_2021}
Maksimovic, M., Walsh, A.~P., Pierrard, V., Štverák, S., \& Zouganelis, I.
  2021, in Kappa {Distributions}: {From} {Observational} {Evidences} via
  {Controversial} {Predictions} to a {Consistent} {Theory} of {Nonequilibrium}
  {Plasmas}, ed. M.~Lazar \& H.~Fichtner, Astrophysics and {Space} {Science}
  {Library} (Cham: Springer International Publishing), 39--51,
  \dodoi{10.1007/978-3-030-82623-9_3}

\bibitem[{Maksimovic {et~al.}(2005)Maksimovic, Zouganelis, Chaufray, Issautier,
  Scime, Littleton, Marsch, McComas, Salem, Lin, \&
  Elliott}]{maksimovic_radial_2005}
Maksimovic, M., Zouganelis, I., Chaufray, J.-Y., {et~al.} 2005, Journal of
  Geophysical Research: Space Physics, 110, \dodoi{10.1029/2005JA011119}

\bibitem[{Maksimovic {et~al.}(2020)Maksimovic, Bale, Berčič, Bonnell, Case,
  Wit, Goetz, Halekas, Harvey, Issautier, Kasper, Korreck, Jagarlamudi,
  Lahmiti, Larson, Lecacheux, Livi, MacDowall, Malaspina, Martinović,
  Meyer-Vernet, Moncuquet, Pulupa, Salem, Stevens, Štverák, Velli, \&
  Whittlesey}]{maksimovic_anticorrelation_2020}
Maksimovic, M., Bale, S.~D., Berčič, L., {et~al.} 2020, The Astrophysical
  Journal Supplement Series, 246, 62, \dodoi{10.3847/1538-4365/ab61fc}

\bibitem[{Marsch(2006)}]{marsch_kinetic_2006}
Marsch, E. 2006, Living Reviews in Solar Physics, 3, 1,
  \dodoi{10.12942/lrsp-2006-1}

\bibitem[{Marsch {et~al.}(1989)Marsch, Pilipp, Thieme, \&
  Rosenbauer}]{marsch_cooling_1989}
Marsch, E., Pilipp, W.~G., Thieme, K.~M., \& Rosenbauer, H. 1989, Journal of
  Geophysical Research, 94, 6893, \dodoi{10.1029/JA094iA06p06893}

\bibitem[{Martinovi\'{c} {et~al.}(2022)}]{Martinovic:prep}
Martinovi\'{c}, M., {et~al.} 2022, The Astrophysical Journal, in preparation

\bibitem[{McComas {et~al.}(2007)McComas, Velli, Lewis, Acton, Balat-Pichelin,
  Bothmer, Dirling~Jr., Feldman, Gloeckler, Habbal, Hassler, Mann, Matthaeus,
  McNutt~Jr., Mewaldt, Murphy, Ofman, Sittler~Jr., Smith, \&
  Zurbuchen}]{mccomas_understanding_2007}
McComas, D.~J., Velli, M., Lewis, W.~S., {et~al.} 2007, Reviews of Geophysics,
  45, \dodoi{10.1029/2006RG000195}

\bibitem[{Müller {et~al.}(2020)Müller, Cyr, Zouganelis, Gilbert, Marsden,
  Nieves-Chinchilla, Antonucci, Auchère, Berghmans, Horbury, Howard, Krucker,
  Maksimovic, Owen, Rochus, Rodriguez-Pacheco, Romoli, Solanki, Bruno,
  Carlsson, Fludra, Harra, Hassler, Livi, Louarn, Peter, Schühle, Teriaca,
  Iniesta, Wimmer-Schweingruber, Marsch, Velli, Groof, Walsh, \&
  Williams}]{muller_solar_2020}
Müller, D., Cyr, O. C.~S., Zouganelis, I., {et~al.} 2020, Astronomy \&
  Astrophysics, 642, A1, \dodoi{10.1051/0004-6361/202038467}

\bibitem[{Neugebauer \& Snyder(1962)}]{neugebauer_solar_1962}
Neugebauer, M., \& Snyder, C.~W. 1962, Science, 138, 1095

\bibitem[{Parker(1965)}]{parker_dynamical_1965}
Parker, E. 1965, Space Science Reviews, 4, \dodoi{10.1007/BF00216273}

\bibitem[{Parker(1958)}]{parker_dynamics_1958}
Parker, E.~N. 1958, The Astrophysical Journal, 128, 664, \dodoi{10.1086/146579}

\bibitem[{Parker(1960)}]{parker_hydrodynamic_1960}
---. 1960, The Astrophysical Journal, 132, 821, \dodoi{10.1086/146985}

\bibitem[{Parker(2010)}]{parker_kinetic_2010}
---. 2010, AIP Conference Proceedings, 1216, 3, \dodoi{10.1063/1.3395887}

\bibitem[{Pierrard \& Lemaire(1996)}]{pierrard_lorentzian_1996}
Pierrard, V., \& Lemaire, J. 1996, Journal of Geophysical Research: Space
  Physics, 101, 7923, \dodoi{10.1029/95JA03802}

\bibitem[{Pilipp {et~al.}(1987)Pilipp, Miggenrieder, Montgomery, Mühlhäuser,
  Rosenbauer, \& Schwenn}]{pilipp_characteristics_1987}
Pilipp, W.~G., Miggenrieder, H., Montgomery, M.~D., {et~al.} 1987, Journal of
  Geophysical Research: Space Physics, 92, 1075,
  \dodoi{10.1029/JA092iA02p01075}

\bibitem[{Rosenbauer {et~al.}(1977)Rosenbauer, Schwenn, Marsch, Meyer,
  Miggenrieder, Montgomery, Muehlhaeuser, Pilipp, Voges, \&
  Zink}]{rosenbauer_survey_1977}
Rosenbauer, H., Schwenn, R., Marsch, E., {et~al.} 1977, Journal of Geophysics
  Zeitschrift Geophysik, 42, 561

\bibitem[{Salem {et~al.}(2021)Salem, Pulupa, Bale, \&
  Verscharen}]{salem_precision_2021}
Salem, C.~S., Pulupa, M., Bale, S.~D., \& Verscharen, D. 2021, Precision
  {Electron} {Measurements} in the {Solar} {Wind} at 1 au from {NASA}'s {Wind}
  {Spacecraft}, Tech. Rep. arXiv:2107.08125, arXiv,
  \dodoi{10.48550/arXiv.2107.08125}

\bibitem[{Scudder(1992)}]{scudder_causes_1992}
Scudder, J.~D. 1992, The Astrophysical Journal, 398, 299,
  \dodoi{10.1086/171858}

\bibitem[{Whittlesey {et~al.}(2020)Whittlesey, Larson, Kasper, Halekas,
  Abatcha, Abiad, Berthomier, Case, Chen, Curtis, Dalton, Klein, Korreck, Livi,
  Ludlam, Marckwordt, Rahmati, Robinson, Slagle, Stevens, Tiu, \&
  Verniero}]{whittlesey_solar_2020}
Whittlesey, P.~L., Larson, D.~E., Kasper, J.~C., {et~al.} 2020, The
  Astrophysical Journal Supplement Series, 246, 74,
  \dodoi{10.3847/1538-4365/ab7370}

\bibitem[{Zouganelis {et~al.}(2004)Zouganelis, Maksimovic, Meyer-Vernet, Lamy,
  \& Issautier}]{zouganelis_transonic_2004}
Zouganelis, I., Maksimovic, M., Meyer-Vernet, N., Lamy, H., \& Issautier, K.
  2004, The Astrophysical Journal, 606, 542, \dodoi{10.1086/382866}

\end{thebibliography}

\end{document}